\documentclass[runningheads]{llncs}
\usepackage[T1]{fontenc}
\usepackage{hyperref}
\usepackage{cleveref}
\usepackage{graphicx}

\begin{document}

\title{An Approach to Evaluate User Interfaces in a Scholarly Knowledge Communication Domain}
\titlerunning{An approach to evaluate user interfaces for knowledge communication}

\author{
	Denis Obrezkov\orcidID{0000-0001-8822-2932} \and
	Allard Oelen\orcidID{0000-0001-9924-9153} \and
	S\"oren Auer\orcidID{0000-0002-0698-2864}}

\authorrunning{Obrezkov et al.}

\institute{TIB Leibniz Information Centre for Science and Technology, Hannover, Germany\\
	\email{\{denis.obrezkov,allard.oelen,soeren.auer\}@tib.eu}}

\maketitle

\begin{abstract}
The amount of research articles produced every day is overwhelming: scholarly knowledge is getting harder to communicate and easier to get lost. A possible solution is to represent the information in knowledge graphs: structures representing knowledge in networks of entities, their semantic types, and relationships between them. But this solution has its own drawback: given its very specific task, it requires new methods for designing and evaluating user interfaces.
In this paper, we propose an approach for user interface evaluation in the knowledge communication domain. We base our  methodology on the well-established  Cognitive Walkthough approach but employ a different set of questions, tailoring the method towards domain-specific needs. We demonstrate our approach on a scholarly knowledge graph implementation called Open Research Knowledge Graph~(ORKG).
\end{abstract}

\section{Introduction}
Modern researchers face numerous problems while conducting research: it is time-consuming to find information, cumbersome to get overviews of related work, and difficult to communicate their results to the right audience. Even though nowadays scholarly articles are often available digitally in PDF form on the Internet, the overwhelming quantity of these unstructured text documents makes it difficult for new knowledge to be discovered, crystallized, and used. 

One approach to address the above-mentioned challenges is to utilize knowledge graphs. Knowledge graphs allow to communicate the actual knowledge of research and provide an alternative to the existing format of a narrative paper-based scholarly communication. Although several well-established implementations of knowledge graphs exist, such as Wikidata~\cite{Vrandecic2014} and DBpedia~\cite{auer2007dbpedia}, there is no such widely-adopted solution in the scholarly knowledge communication domain. In order to investigate this phenomenon and to improve knowledge communication from and to researchers, it is promising to analyze how information is actually transferred between a researcher and a knowledge graph interface.

In this paper, we propose a Cognitive Walkthrough methodology that can be employed to identify issues in user interfaces for scholarly knowledge communication. In \autoref{section:related-work} we describe knowledge graphs and 
how can we adopt an interface evaluation tool to account for domain-specific usability issues. In \autoref{section:methodology}, we describe the resulting methodology of our walkthrough. 

\section{Related work}
\label{section:related-work}

A promising approach to organizing scholarly knowledge is using knowledge graphs. Similar to graph databases, knowledge graphs consist of entity networks and their respective relations. Additionally, knowledge graphs include semantics~\cite{kroetsch2016special}, often represented using ontologies, to capture the meaning of the data.

\begin{figure}
    \centering
    \includegraphics[width=.8\textwidth]{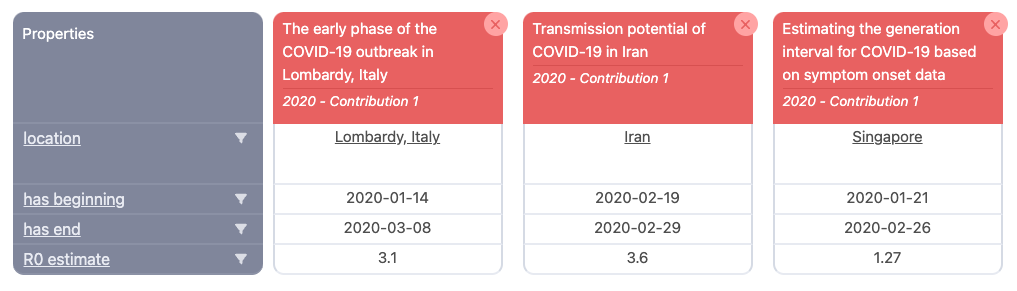}
    \caption{Excerpt of an ORKG comparison with the columns representing papers.}
    \label{fig:orkg-comparison}  
\end{figure}

A small number of approaches exist to represent scholarly knowledge using knowledge graphs. Among others, this includes the Semantic Scholar Academic Graph~\cite{wade2022semantic}, Microsoft Academic Graph~\cite{herrmannova2016analysis} and the Open Research Knowledge Graph (ORKG)~\cite{auer2020improving}. The former two approaches focus mainly on representing bibliographic metadata, while the latter focuses on describing and representing the actual knowledge stated in scientific articles, additionally allowing tabular literature overviews, called Comparisons~\cite{Oelen2020}. A typical Comparison lists several chosen properties for a number of papers, thus, allowing a researcher to identify the important concepts of the works~(see \autoref{fig:orkg-comparison}, for an example Comparison).

To increase usability of a knowledge communication interface, several well-established methods can be utilized: interface walkthroughs~\cite{nielsen1990heuristic,wharton1994cognitive}, user studies~\cite{hornbaek2006current}, and GOMS-based user modelling~\cite{Card1983ThePO}. A Cognitive Walkthrough~(CW) is an task-centered interface walkthrough method that accounts for user's mental processes and goals~\cite{wharton1994cognitive}. 
The evaluator is asked to perform a number of steps. First, to identify the target audience and their background. Second, to understand their tasks and goals. Third, to specify a correct sequence of actions for each task. Fourth, to perform analysis by answering a set of  predefined questions.
Last, to write down problems and provide success and failure scenarios.

A prominent model to rely on while designing the human-information interaction is Model Human Processor introduced~(MHP).~\cite{Card1983ThePO}. This model represents human mind as an information-processing system and introduces several limitations that should be considered when designing a usable system. First, the working memory capacity, that can store only a limited amount of information, estimated to be around 5 to 7 chunks~\cite{miller1956magical,sousa2016brain}.
Second, the involvement of long-term memory in information processing. It is known that humans understand and memorize information better if the information makes sense for them~\cite{sousa2016brain}. 

To represent user knowledge about the system, a notion of mental model is used. The latter is defined as a set of user beliefs about the system, constructed through the interaction~\cite{norman1983some}. To refer to a user's mental representation of the current state of a digital environment, we use term ``cognitive context''.

\section{Methodology}
\label{section:methodology}

The initial steps in our methodology are similar to the original Cognitive Walkthrough. First, the target audience and their background should be identified. Second, a user goal should be specified, and user tasks should be written down. It should be noted, that we do not require a correct sequence of actions for each task since it can interfere with the evaluation process. In the third step, an evaluator is expected to perform analysis by answering a set of questions from~\autoref{cw-questions}. 
\begin{table}
	\centering
	\caption{A question set for the developed Cogntive Walkthrough.}
	\begin{tabular}{||p{0.05\textwidth}p{0.78\textwidth}p{0.14\textwidth}||}
		\hline
		\textbf{№} &  \textbf{Walkthrough Question} & \textbf{Reference} \\ \hline 

		\textbf{Q1} &  Will users be aware of the steps they have to perform to complete a core task? & \cite{clark2007usability,rizzo1997avanti,wharton1994cognitive} \\ \hline
		\textbf{Q2} & Will users be able to determine how to perform these steps? & \cite{clark2007usability,wharton1994cognitive} \\ \hline
		\textbf{Q3} &  Will users be aware of the application’s status at all times?& \cite{nielsen1990heuristic,wharton1994cognitive} \\ \hline
		\textbf{Q4} &  Will users receive feedback in the same place and modality as where they have performed their action? &  \cite{nielsen1990heuristic,rizzo1997avanti} \\ \hline
		\textbf{Q5} & Will users be able to recognize, and recover from non-critical errors? & \cite{clark2007usability,nielsen1990heuristic} \\ \hline
		\textbf{Q6} & Will users be able to avoid making dangerous errors from which they cannot recover?& \cite{clark2007usability,nielsen1990heuristic} \\ \hline
		\textbf{Q7} & Will users be able to efficiently work with the system, considering the limitations of working memory? & \cite{Card1983ThePO,nielsen1990heuristic} \\ \hline
		\textbf{Q8} & Will users be able to understand the information provided by the system, given their background? & \cite{Card1983ThePO,nielsen1990heuristic} \\ \hline 
	\end{tabular}

	\label{cw-questions}
\end{table}
The listed questions address a user's need to achieve a certain goal~(Q1, Q2), the integrity of a cognitive context~(Q3-Q6), the ease of the presented information processing~(Q7, Q8).
The last step of the walkthrough includes writing down problems and providing failure scenarios.

To demonstrate the methodology, we  perform the developed Cognitive Walkthrough on ORKG. For that purpose, we choose the most prominent task of populating a knowledge graph. 
In our case, a typical user is an early career researcher. The user has a good understanding of her professional scientific field, with occasional knowledge gaps regarding specific methods or concepts. The user's goal is to gain information about concepts related to her own discipline either via structuring new knowledge or exploring structures created by others.

\textit{\textbf{Task: To extract information from papers and represent it with the concepts in a knowledge graph.}}
When a user tries to represent information from a paper, she is provided with a list of suggested properties and has the possibility to choose a template. When choosing a template, the user is given a form with multiple properties to fill. Since it is required to type a query to find the needed template, the user might be confused by not knowing the names of existing templates (Q2, Q8). And since templates are chosen by name and have no preview, she might choose an improper template with an appealing name, e.g. confusing the mental fatigue template with one related to material fatigue~(Q4). 

\begin{figure}[]
    \centering
    \includegraphics[width=1\textwidth]{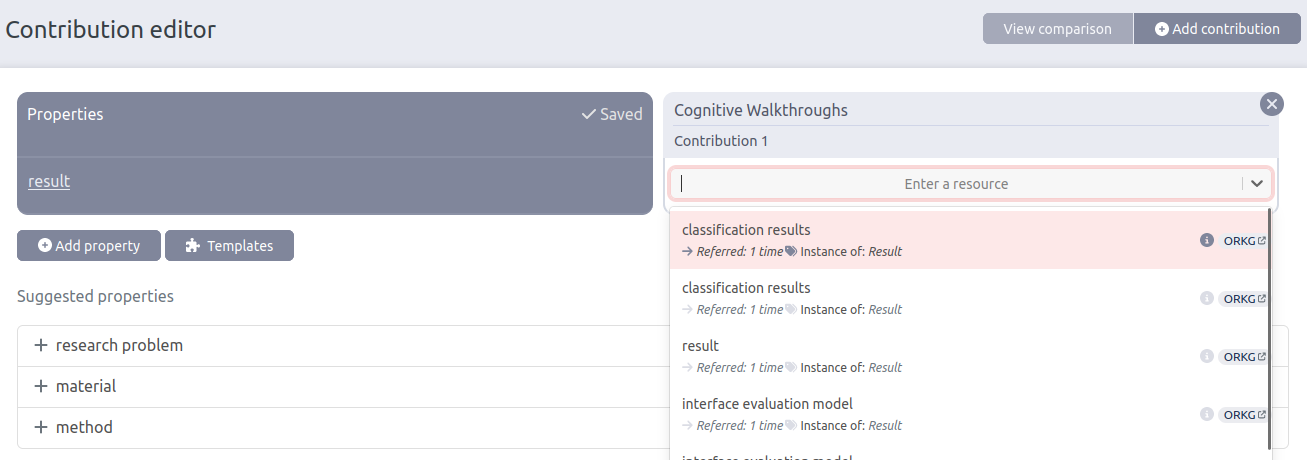}
    \caption{The interface of ORKG contribution editor with several issues.}
    \label{fig:orkg-contribution-editor}
   
\end{figure}

Another problem arises when the user tries to add a property to the contribution. In case a user chooses a property from a template, she is asked to fill in the value of the property. For example, 
when the user chooses the property ``result'' and is asked to ``Enter a resource''~(see~\autoref{fig:orkg-contribution-editor}), it is not clear what the resource is (Q8) and how it affects her initial goal (Q1). At the same time, if a user deletes a property value, there is no easy way to undo the deletion (Q5).

\section{Discussion and Conclusion}
\label{section:discussion}

In this paper, we introduced a new methodology of Cognitive Walkthrough for the knowledge communication domain. We developed our approach based on the principles of the original Cognitive Walkthrough and adopted the notion of cognitive context together with constraints from the Model Human Processor for the evaluation questions. We performed a walkthrough evaluation on a scholarly knowledge graph implementation called ORKG.

During the evaluation, we discovered several issues in the ORKG interface. For example, we observed that the evaluated interface has issues associated with questions Q4 and Q5, thereby breaking the user's context integrity. Given the knowledge about the disturbed cognitive process, we can leverage the appropriate techniques to address the problem, e.g. by utilizing context switching~\cite{gauselmann2023relief} or cognitive offloading~\cite{RISKO2016676}. In our future work we aim to elaborate on connections between walkthrough questions and underlying cognitive processes and to provide the appropriate mitigating techniques for revealed issues.

\textbf{Acknowledgements.} This work was co-funded by the European Research Council for the project ScienceGRAPH (Grant agreement ID: 819536) and the TIB Leibniz Information Centre for Science and Technology.

\newpage
\bibliographystyle{acm}
\bibliography{CWonORKG}

\end{document}